\documentclass[aps,prd,nofootinbib,twocolumn,superscriptaddress,floatfix,notitlepage]{revtex4-1}

\usepackage[dvipsnames]{xcolor}
\usepackage{graphicx}
\usepackage{amsmath,amsfonts,amssymb}
\usepackage[breaklinks,colorlinks,urlcolor=Plum,citecolor=MidnightBlue,linkcolor=WildStrawberry]{hyperref}
\usepackage{enumitem}
\usepackage{orcidlink}
\usepackage{array}
\usepackage{tabularx}

\newcommand{\td}[0]{\mathrm{d}}

\newcommand{\papertitle}{Beam tube boundary effects in stray light modeling of long Fabry–Perot arm cavities for third-generation gravitational-wave detectors}
\begin{document}

\title[]{\papertitle}

\author{M.~Andr\'es-Carcasona\orcidlink{0000-0002-8738-1672}}
\email{mandresc@mit.edu (corresponding author)}
\affiliation{LIGO Laboratory, Massachusetts Institute of Technology, Cambridge, MA 02139, USA}
\author{M.~Evans\orcidlink{0000-0002-8738-1672}}
\affiliation{LIGO Laboratory, Massachusetts Institute of Technology, Cambridge, MA 02139, USA}

\begin{abstract}
Next-generation gravitational-wave detectors such as Cosmic Explorer and the Einstein Telescope will operate 10--40~km Fabry--Perot arm cavities inside vacuum beam tubes. FFT-based paraxial tools treat propagation in free space and therefore do not explicitly enforce beam tube boundary conditions. We introduce a waveguide-like mode description of the optical field that incorporates an imposed beam tube boundary condition and enables an independent benchmark of free-space FFT tools We derive the associated modal-mixing matrices for mirrors and baffles, including a closed-form series for axisymmetric circular apertures. We quantify the strain-equivalent couplings from baffle miscentering and from a localized near-wall tube defect, and show that they are suppressed as baffle density increases. In the relevant regime of densely baffled cavities and small perturbations, beam tube boundary effects are subdominant, which supports the continued use of FFT-based codes to guide the design of 3G detectors.
\end{abstract}

\maketitle

\section{Introduction}

Scattered light, also known as stray light, arises from photons that deviate from the intended path due to scattering off surfaces, and is a critical noise source in gravitational wave (GW) detectors~\cite{Accadia:2010zzb,Fiori:2020arj,Was:2020ziy,LIGO:2020zwl,Longo:2020onu,Longo:2021avq,Longo:2023vac,Andres-Carcasona:2024hel,Macquet:2022simsVirgo,Andres-Carcasona:2022imx,Romero-Rodriguez:2020bys,Romero-Rodriguez:2022mje,Ballester:2021bua,Bianchi:2021unp}. Such stray light can recombine with the resonant laser field, introducing both phase and amplitude noise that may significantly degrade the detector sensitivity. This effect is particularly problematic because it can couple to the mechanical motion of non-optical components, resulting in low-frequency noise artifacts that are difficult to model or mitigate using standard isolation techniques~\cite{Ottaway:2012oce}.

To control scattered light, GW detectors employ baffles and beam dumps to absorb or redirect unwanted photons. These components are carefully manufactured so that the reflection and scattering of light is reduced. For next-generation detectors, with longer propagation baselines and higher expected sensitivity, the design must ensure that scattered light does not limit the detector performance. In the absence of full-scale prototypes, the design and selection of stray light mitigation strategies must be guided primarily by modeling. This implies that modeling tools must be systematically verified and validated by benchmarking between independent numerical approaches, analytic limits, and, where available, experimental data from current detectors or dedicated test measurements.

Current approaches to scattered-light modeling in GW detectors rely mostly on ray-tracing simulations, fast Fourier transform (FFT) codes, and simplified analytical approximations, combined with empirical measurements~\cite{LIGO:2020zwl,Andres-Carcasona:2023qom}. Ray-tracing methods can accommodate very complex geometries, at the expense of substantial computing resources, as they must simulate a sufficiently large number of photon trajectories to extract statistically significant conclusions. By contrast, FFT-based codes can compute the field at an arbitrary position along these long FP cavities, but complex geometries cannot be included, as they rely on propagating the field in free space under the paraxial approximation. One of these FFT codes is the Stationary Interferometer Simulation (SIS) package~\cite{Romero21}, which has been tested against modal-expansion methods and compared with photodiode measurements in LIGO and with an instrumented baffle in Virgo~\cite{Vajente:22,VajenteCEbaffles,Andres-Carcasona:2022imx,Romero-Rodriguez:2020bys,Romero-Rodriguez:2022mje,Ballester:2021bua}. 

Third-generation (3G) GW detectors, such as the Einstein Telescope (ET)~\cite{ETcds,ETdesign} and Cosmic Explorer (CE)~\cite{CE1,CE2,CE3}, will require very long FP cavities of 10~km and 40~km, respectively. Longer baselines and tighter noise budgets increase the sensitivity to stray-light noise, requiring more accurate modeling to guide the design. In the main FP cavities, the primary tool currently used to inform the baffle layout is SIS for both observatories~\cite{Andres-Carcasona:2023qom,Andres-Carcasona:2025xwq,BTB_preliminary}. An open question is whether, for the larger baffle separations expected in 3G designs, interactions with the surrounding vacuum beam tube remain negligible. This point is particularly relevant because FFT-based propagation does not impose beam tube boundary conditions and if wall interactions are non-negligible, SIS predictions may be inaccurate.

In this work we provide an independent benchmark of SIS in the specific regime relevant to 3G arm-cavity baffle design. Our approach introduces a waveguide-like modal description that enforces an imposed beam tube boundary condition while remaining computationally tractable over 40~km propagation lengths. This enables a direct comparison of steady-state intracavity fields between a model that is aware of the beam tube and the free-space one, and a systematic assessment of when beam tube interactions can be safely neglected. We find that, for baffled cavities with CE-like parameters, the beam tube boundary primarily affects the large-radius tail of the field, while the resonant field within the mirror aperture and the noise couplings relevant for small baffle motion are largely unchanged. Additionally, we show that baffles inside the cavity act as spatial filters that progressively suppress the diffractive halo and reduce the field incident on the beam tube, making this boundary effect less relevant. As a result, SIS remains a reliable tool for guiding baffle layout studies in the densely baffled, small-perturbation regime targeted by current 3G designs.

Modal decomposition techniques for scattered-light studies have been explored in the literature. For example, Refs.~\cite{VajenteCEbaffles,Vajente:22} use Hermite--Gauss (HG) and Laguerre--Gauss (LG) modes to model stray light generated by finite baffle apertures within the paraxial, free-space context. In contrast, the waveguide-like basis introduced here incorporates an imposed beam tube boundary condition, allowing a direct benchmark against free-space FFT propagation in the regime of long baselines relevant to 3G detectors.

The remainder of this paper is organized as follows. In Sec.~\ref{sec:newmodes}, we present a summary of set of modes employed in this work. Section~\ref{sec:steady_state} explains how to obtain the steady-state field and how noise couplings can be computed. Finally, in Sec.~\ref{sec:results}, we perform numerical experiments with a CE-like cavity to evaluate the limitations of current FFT-based tools. The mathematical details and some numerical considerations are presented in the Appendix.

\section{Waveguide-like modes} \label{sec:newmodes}
We consider the propagation of a collimated and coherent laser beam in vacuum and work in the monochromatic regime. Under a fixed polarization, the field satisfies the scalar Helmholtz equation:
\begin{equation}\label{eq:Helmholtz}
    \nabla^2\psi + k^2\psi = 0\, ,
\end{equation}
where $\psi$ is the complex valued function that describes the field and $k$ is the wavenumber and is related to the wavelength of the laser $\lambda$ as $k=2\pi/\lambda$. In many applications, the scales in the beam propagation direction are much bigger than those in the transverse plane, and the paraxial approximation is used~\cite{siegman1986lasers}. Under this approximation and vacuum conditions the field is usually well-described by a Gaussian beam. This beam is characterized by a Gaussian transverse intensity distribution and by an evolution that smoothly varies during its propagation. The field can be expressed as
\begin{multline}\label{eq:gaussian_beam}
    \psi(x,y,z) = \sqrt{\frac{2}{\pi w(z)^2}} 
\exp\!\Biggl[
  -ikz \\ -ik\frac{x^2+y^2}{2R(z)} + i\varphi_g(z) - \frac{x^2+y^2}{w(z)^2}
\Biggr]\, ,
\end{multline}
where the beam size $w(z)$, the wavefront radius of curvature $R(z)$, and the Gouy phase $\varphi_g(z)$ are given by
\begin{align}
w(z) &= w_0 \sqrt{1 + \left( \frac{z}{z_R} \right)^2}, \label{eq:wz} \\[4pt]
R(z) &= z \left[ 1 + \left( \frac{z_R}{z} \right)^2 \right], \label{eq:Rz} \\[4pt]
\varphi_g(z) &= \arctan\!\left(\frac{z}{z_R}\right), \label{eq:gouy}
\end{align}
with the Rayleigh range defined as $z_R = \pi w_0^2/\lambda$ and $w_0$ denoting the beam waist size, reached at $z=0$.

In real-world experiments, beams are not exactly Gaussian, and the field is usually described by a mode expansion using a set of eigenvectors of the paraxial propagation equation. The two sets of eigenmodes that are typically used are the HG and LG as they are very efficient in FP cavities like the ones used in GW experiments~\cite{freise2013}. These mode bases can also be used to model the clipping by finite apertures, such as baffles, and mirrors, though a large number of modes must be tracked to achieve sufficient resolution ~\cite{kogelnik1964coupling,Tanaka72,Vajente:22,VajenteCEbaffles,Petrovic2002}. Unfortunately, these modes do not account for the boundary condition that the beam tube that hosts this laser beam imposes on the field.

We therefore use a set of modes that include hard-wall boundary condition. This is done by solving Eq.~\eqref{eq:Helmholtz} in cylindrical coordinates and without the need of relying on the paraxial approximation. The boundary conditions are 0 at the beam tube radius (denoted as $R$) and periodic in the polar angle: $\psi(r=R,\phi,z) = 0$ and $\psi(r,\phi,z)=\psi(r,\phi+2\pi,z)$. The derivation of the generic set of modes is presented in Appendix~\ref{sec:newmodes}. Nonetheless, for all the numerical experiments that we carry out, there is a symmetry under the transformation $\phi\to -\phi$, so we can use as a basis the modes
\begin{equation}
     \psi_{mn}(r,\phi,z) = J_m\left(\frac{\alpha_{mn}r}{R}\right)\cos(m\phi)e^{-i\beta_{mn}z}\, ,
\end{equation}
reducing the number of modes to keep track of and the computational cost of the numerical experiments. Here, $m\geq 0$, $n\in\mathbb{N}$, $\alpha_{mn}$ represents the $n^{\mathrm{th}}$ root of the $J_m$ Bessel function and 
\begin{equation}
    \beta_{mn} = \sqrt{k^2-\left( \frac{\alpha_{mn}}{R}\right)^2}
\end{equation}

\begin{figure}[htbp]
    \centering
    \includegraphics[width=1.0\linewidth]{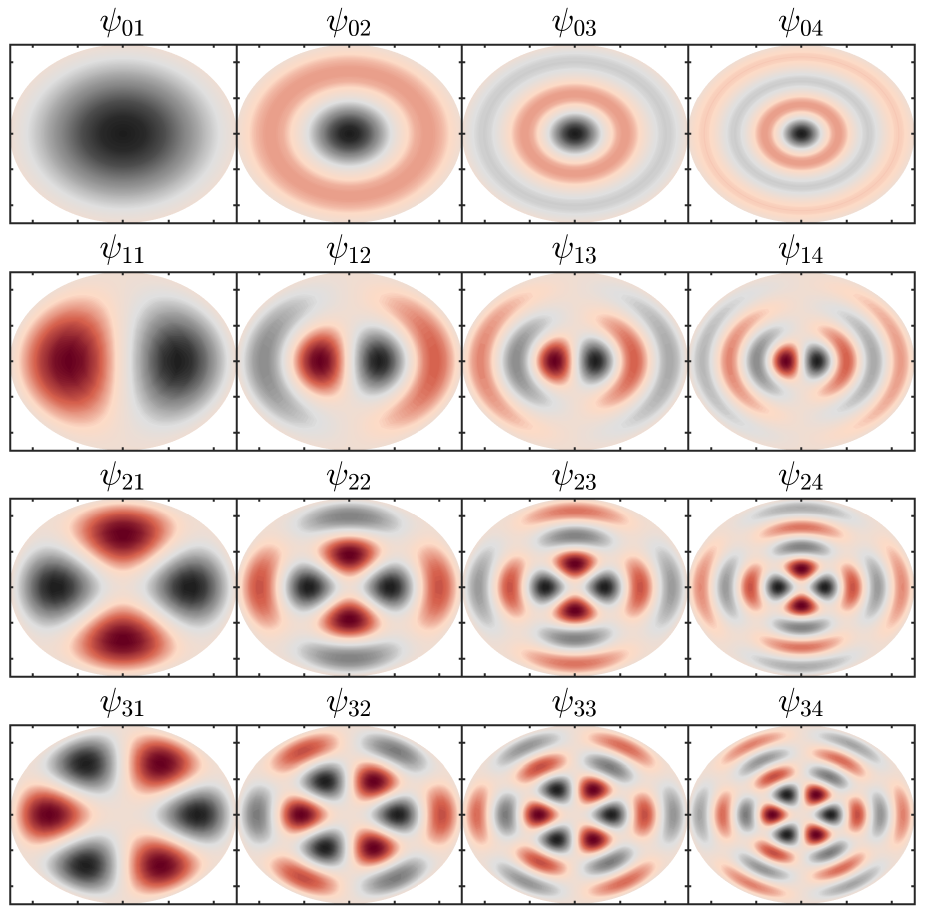}
    \caption{Plot of the first $\psi_{mn}$ modes. Arbitrary color scale.}
    \label{fig:psimn_plot}
\end{figure}

In Fig.~\ref{fig:psimn_plot} we show the sixteen lowest-order eigenmodes $\psi_{mn}$ of this new basis. 
Modes with $m=0$ are azimuthally symmetric, exhibiting a purely radial structure. 
For $m>0$, the angular dependence produces $2m$ lobes arranged uniformly in the azimuthal direction. 
The radial index $n$ determines the number of nodes in the radial coordinate as increasing $n$ introduces additional concentric rings, corresponding to finer spatial structure in the mode profile. 
Higher-order modes display progressively richer angular patterns and smaller-scale radial features.

Once the field is expressed in the $\{\psi_{mn}\}$ basis, the analysis reduces to evolving the vector of modal coefficients. Free propagation with these modes and assumptions over an arbitrary distance does not mix modes and therefore the operator is a diagonal matrix that modifies the phase of each mode according to $\beta_{mn}$. Any optical element located in a transverse plane, like mirrors, baffles, and more general phase or amplitude masks, is modeled as a multiplicative transverse operator $Q(r,\phi)$ acting on the field. In the modal representation, such operators generally induce mode mixing, with coupling coefficients given by overlap integrals between the modes weighted by $Q(r,\phi)$. This provides a unified framework in which diffraction from finite apertures and other transverse inhomogeneities are captured through such mode-mixing matrices, while propagation remains diagonal and inexpensive. The full construction of the propagation and mode-mixing matrices used throughout this work are derived in Appendix~\ref{sec:modal_mixing} with an analytical closed form for some of them. 

The notation that we employ is $\mathbf{Q}$ for the mode-mixing matrix of a mirror (while $\mathbf{R}$ and $\mathbf{T}$ indicate the corresponding reflection and transmission operators, respectively), $\mathbf{S}(d_x)$ for that of a baffle that may be miscentered along the $x$ axis by an amount $d_x$, and $\mathbf{P}(\Delta z)$ for the propagation operator over a distance $\Delta z$. A tilde on these operators indicates that the explicit $z$-dependent propagation phase has been factored out (see Appendix~\ref{sec:modal_mixing} for the details).

\section{Steady state field and noise}\label{sec:steady_state}
\begin{figure*}[htbp]
    \centering
    \includegraphics[width=1.0\linewidth]{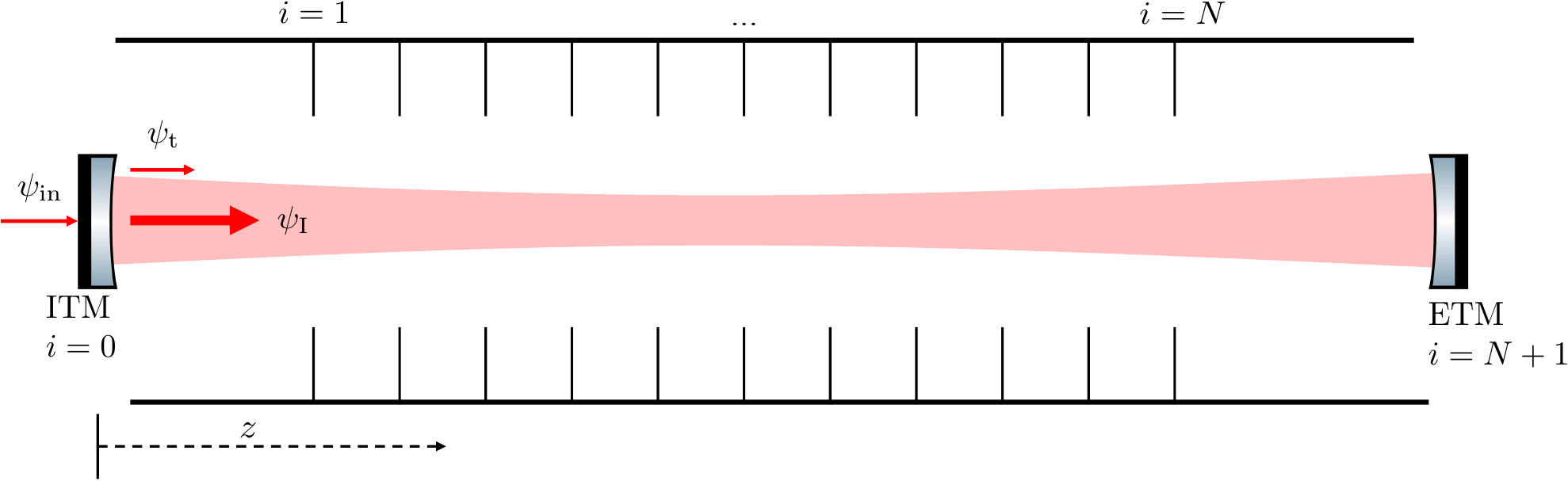}
    \caption{Schematic of a baffled Fabry–Pérot cavity showing the relevant fields and indices.}
    \label{fig:ScxhemeWGpaper}
\end{figure*}

In order to study the properties of the FP cavity, the steady state field is usually required. To construct the round-trip operator, we start from the intracavity field outgoing from the input test mass (ITM), denoted $\psi_{\mathrm{I}}$. This field propagates a distance $L$ along the cavity to reach the end test mass (ETM), where it is reflected, and then propagates back the same distance to the ITM. Then, the field is reflected at the ITM to begin strat the next pass. This sequence of operations is illustrated schematically in Fig.~\ref{fig:ScxhemeWGpaper}. In the absence of baffles, the round-trip matrix is simply
\begin{equation}\label{eq:ShortA}
    \mathbf{A} = \mathbf{R}_{\mathrm{ITM}}\mathbf{R}_{\mathrm{ETM}}\, .
\end{equation}

The simplicity of this expression comes from the fact that the reflection matrices, $\mathbf{R}_{\mathrm{ITM}}$ and $\mathbf{R}_{\mathrm{ETM}}$, already account for the propagation of the field, as they are referred to a given $z$-plane:
\begin{equation}
    \mathbf{R}_{\rm ITM} = r_{\rm ITM} \mathbf{Q}(z_{\rm ITM})\,, \quad \,\mathbf{R}_{\rm ETM} = r_{\rm ETM} \mathbf{Q}(z_{\rm ETM})\,.
\end{equation}

If one desires to separate the reflection and the propagation operators, then the round-trip matrix can be written as
\begin{equation}\label{eq:FullA}
    \mathbf{A} = \tilde{\mathbf{R}}_{\mathrm{ITM}}\,\mathbf{P}(-L)\,\tilde{\mathbf{R}}_{\mathrm{ETM}}\,\mathbf{P}(L)\, ,
\end{equation}
where
\begin{equation}
    \tilde{\mathbf{R}}_{\rm ITM} = r_{\rm ITM} \tilde{\mathbf{Q}}(z_{\rm ITM})\,, \quad \,\tilde{\mathbf{R}}_{\rm ETM} = r_{\rm ETM} \tilde{\mathbf{Q}}(z_{\rm ETM})\,.
\end{equation}

Note how Eq.~\eqref{eq:ShortA} and Eq.~\eqref{eq:FullA} are completely equivalent, but the latter makes explicit the various steps involved. In fact, Eq.~\eqref{eq:FullA} is the same quoted in Ref.~\cite{VajenteCEbaffles}.

The intracavity field also receives a continuous contribution from the injected beam transmitted through the ITM. Denoting this transmitted component by $\psi_t = \mathbf{T}_{\mathrm{ITM}}\psi_{\mathrm{in}}$, the self-consistency relation for the intracavity field reads
\begin{equation}
    \psi_{\mathrm{I}} = \mathbf{A}\,\psi_{\mathrm{I}} + \psi_t \, ,
\end{equation}
which after rearranging the terms, yields the equation for the steady-state solution:
\begin{equation}
    (\mathbf{I}-\mathbf{A})\,\psi_{\mathrm{I}} = \mathbf{T}_{\mathrm{ITM}}\,\psi_{\mathrm{in}}\, .
\end{equation}
Expressed in terms of the vector of modal expansion coefficients of the intracavity and input fields, $\mathbf{c}_{\mathrm{I}}$ and $\mathbf{c}_{\mathrm{in}}$, this equation becomes
\begin{equation} \label{eq:cIcin}
    (\mathbf{I}-\mathbf{A})\,\mathbf{c}_{\mathrm{I}} = \mathbf{T}_{\mathrm{ITM}}\,\mathbf{c}_{\mathrm{in}}\, .
\end{equation}

When baffles are present, the round-trip matrix must be modified to include their clipping action and the propagation between successive baffles. In this case, the operator reads as~\cite{VajenteCEbaffles}
\begin{multline}
    \mathbf{A} = \tilde{\mathbf{R}}_{\mathrm{ITM}}
    \left[ \prod_{i=N}^{1} \tilde{\mathbf{S}}(z_i|d_x)\,\mathbf{P}(z_{i}-z_{i+1}) \right] \\ \times
    \tilde{\mathbf{R}}_{\mathrm{ETM}}
    \left[ \prod_{i=1}^{N} \tilde{\mathbf{S}}(z_i|d_x)\,\mathbf{P}(z_i-z_{i-1}) \right] \, .
\end{multline}



Additionally, we allow for the baffles to have a slight transverse offset $d_x$ from the beam tube axis, which introduces an asymmetry in the mode coupling and, as discussed below, leads to an additional phase shift of the intracavity field after a full round trip. This phase shift is the origin of the scattered-light noise associated with baffle motion.

To quantify this effect, we work directly with the steady-state intracavity field on a transverse grid. For a given cavity configuration we compute the complex field at the ITM plane, $\psi_0(x,y)$, for the nominal (centred) baffle positions. We then repeat the calculation with one baffle displaced by a small offset $d_x$, obtaining a second field $\psi_{d_x}(x,y)$ at the same plane. In both cases we are interested in the phase of the fundamental cavity mode, which we approximate by a prescribed fundamental Gaussian beam $u_{00}(x,y)$.

We extract the complex amplitude of the fundamental mode from each field by taking the normalized projection onto $u_{00}$ 
\begin{equation}
    a_0 = \frac{\langle u_{00}, \psi_0 \rangle}{\langle u_{00}, u_{00} \rangle}\,, 
    \qquad
    a_{d_x} = \frac{\langle u_{00}, \psi_{d_x} \rangle}{\langle u_{00}, u_{00} \rangle}\,.
\end{equation}

The induced phase shift of the fundamental component at the ITM plane when the baffle is displaced is then obtained from the relative phase of these coefficients,
\begin{equation}
    \Delta\phi(d_x) = \arg\!\left(\frac{a_{d_x}}{a_0}\right)\,,
\end{equation}
and this phase shift can be interpreted as an effective change in the cavity length. For a FP cavity, the round-trip phase of the fundamental mode is $\phi = 2kL$, so a small phase change $\Delta\phi$ corresponds to an equivalent longitudinal displacement
\begin{equation}
    \Delta L = \frac{\Delta\phi}{2k}\,.
\end{equation}
Dividing by the arm length we obtain the corresponding phase to noise coupling or equivalent strain,
\begin{equation} \label{eq:h}
    h \equiv \frac{\Delta L}{L}
    = \frac{\Delta\phi}{2kL}\,.
\end{equation}

In a full noise budget one would combine this with the baffle displacement spectrum, but here we use Eq.~\eqref{eq:h} as an effective coupling metric to compare various configurations.

\section{Results}
\label{sec:results}
In this section we present a set of results and examples that illustrate and validate the formalism introduced in this work. 
All the expressions have been implemented in \texttt{Python} and the analytical clipping matrix requires the \texttt{mpmath} library, which allows for arbitrary-precision arithmetic operations. The use of arbitrary precision is essential in this context, as the computation of factorials and special functions can involve extremely large intermediate values that would otherwise lead to numerical instabilities or loss of accuracy.

All the numerical examples use the CE-representative values displayed in Tab.~\ref{tab:ParamsTable} unless otherwise stated. 

\newcommand{\desc}[1]{\parbox[t]{4.5cm}{\raggedright #1}}

\begin{table}[htbp]
\centering
\small
\setlength{\tabcolsep}{3pt}
\renewcommand{\arraystretch}{1.05}
\begin{tabular}{l c c l}
\hline\hline
\textbf{Name} & \textbf{Value} & \textbf{Units} & \textbf{Description} \\
\hline
$w_0$              & 6.9       & cm  & \desc{Gaussian beam waist} \\
$L$                & 40        & km  & \desc{Fabry--Pérot cavity length} \\
$\lambda$          & 1064      & nm  & \desc{Laser wavelength} \\
$R$                & 0.60       & m   & \desc{beam tube inner radius} \\
$R_b$              & 0.50      & m   & \desc{Baffle inner radius} \\
$R_m$              & 0.375     & m   & \desc{Mirror clear aperture radius} \\
$r_{\mathrm{ITM}}$ & 0.9930    & --  & \desc{ITM scalar amplitude reflectivity} \\
$t_{\mathrm{ITM}}$ & 0.1183    & --  & \desc{ITM scalar amplitude transmissivity} \\
$r_{\mathrm{ETM}}$ & 0.9999975 & --  & \desc{ETM scalar amplitude reflectivity} \\
\hline\hline
\end{tabular}
\caption{Parameters used for the examples unless stated otherwise.}
\label{tab:ParamsTable}
\end{table}

\subsection{Comparison of analytical and numerical baffle clipping matrix}
We can begin by comparing the analytical expression derived for the clipping of the baffle and the result that one would obtain using a numerical integration. This will validate Eq.~\eqref{eq:Smnpq}.

\begin{figure*}[htbp]
    \centering
    \includegraphics[width=1.0
    \linewidth]{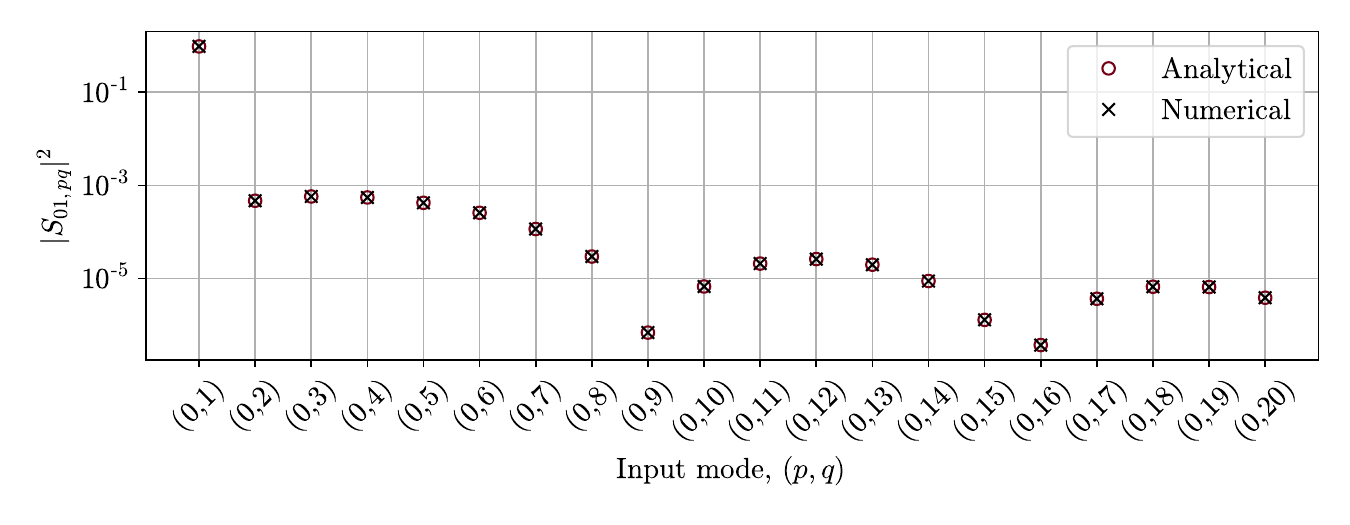}
    \caption{Comparison of the coupling matrix calculation using the analytical expression derived and a numerical integration on a uniform grid of $4096\times 4096$ points.}
    \label{fig:Smnpq_comparison}
\end{figure*}

Considering the parameters from Tab.~\ref{tab:ParamsTable}, we compute the values of the clipping matrix of the mode $(m,n)=(0,1)$ to various modes with $p=0$ (as any other value will simply not couple) and for $q\in[1,20]$. For the analytical calculation we truncate the summation in $j$ when the contribution of the term is smaller than a threshold of $10^{-30}$. For the numerical integration we use a uniform grid over the square domain 
$[-R,R]\times[-R,R]$ with $4096$ points in each spatial direction. The results are displayed in Fig.~\ref{fig:Smnpq_comparison}. The average total time to compute these coupling coefficients on a \texttt{AMD EPYC 7502} CPU is $0.69$~s for the analytical approach and $731.4$~s for the numerical one. This implies an average of $0.03$~s per mode for the analytical approach and $36.6$~s for the numerical, which represents a speed-up of more than three orders of magnitude. Nonetheless, the computational speed improvement of the analytical solution is not that significant when one evaluates couplings of even higher order modes, as more terms are required to achieve convergence and with it, higher precision from the factorial and Gamma function are required. To this end, in practice we implement a more efficient numerical method as explained in Appendix~\ref{app:S_numeric}. 

 It is worth noting that the values of the coupling coefficients decrease rapidly with increasing $q$, spanning several orders of magnitude as illustrated in Fig.~\ref{fig:Smnpq_comparison}. 
This behavior reflects the physical intuition that higher-order radial modes overlap less efficiently with the fundamental mode when the clipping region is relatively small. 
In practice, this rapid suppression implies that only a limited number of low-order couplings need to be considered in realistic calculations.

\subsection{Reconstruction and clipping of a Gaussian beam}



Another relevant test is the reconstruction of a Gaussian beam. Since the FP cavities of GW detectors are tuned such that the fundamental Gaussian mode resonates inside of them, it is essential to verify that the waveguide basis is capable of accurately reproducing this field. 

\begin{figure}[htbp]
    \centering
    \includegraphics[width=1\linewidth]{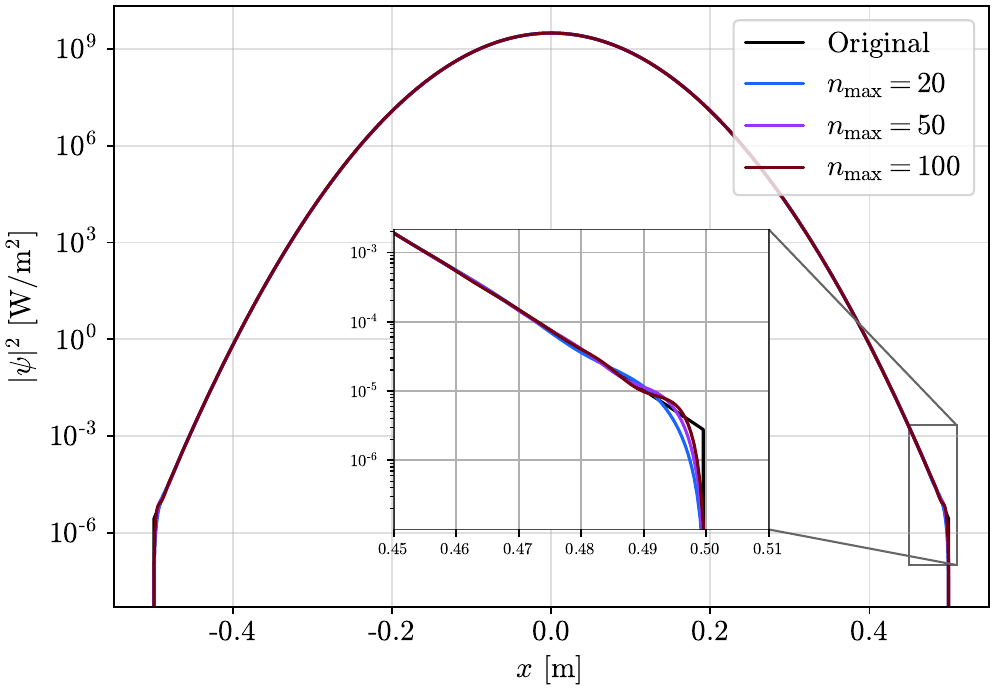}
    \caption{Comparison between a Gaussian beam profile along $y=0$ (black line) and its reconstruction using the waveguide-like modes with azimuthal index $m=0$ and different truncation orders in the radial index, $n_{\max}$.}
    \label{fig:fieldreconstruction_Orig}
\end{figure}

We can start by creating a Gaussian beam defined in the domain that fulfills $x^2+y^2<R^2$ (as our modes cannot describe anything beyond the radius of the beam tube) and try to reconstruct it using different amounts of modes. For this case, only modes with azimuthal index $m=0$ are included, since the cylindrical symmetry of the Gaussian beam ensures that all coefficients with $m\neq 0$ vanish. The result is displayed in Fig.~\ref{fig:fieldreconstruction_Orig}. 

Even though that the lowest waveguide mode is not a Gaussian, we see that a superposition of a few of these modes is already enough to capture correctly the Gaussian beam profile. The higher the number of these modes, the better the accuracy, which in this particular case translates into a better description of the tail near the beam tube edge.

\begin{figure}[htbp]
    \centering
    \includegraphics[width=1\linewidth]{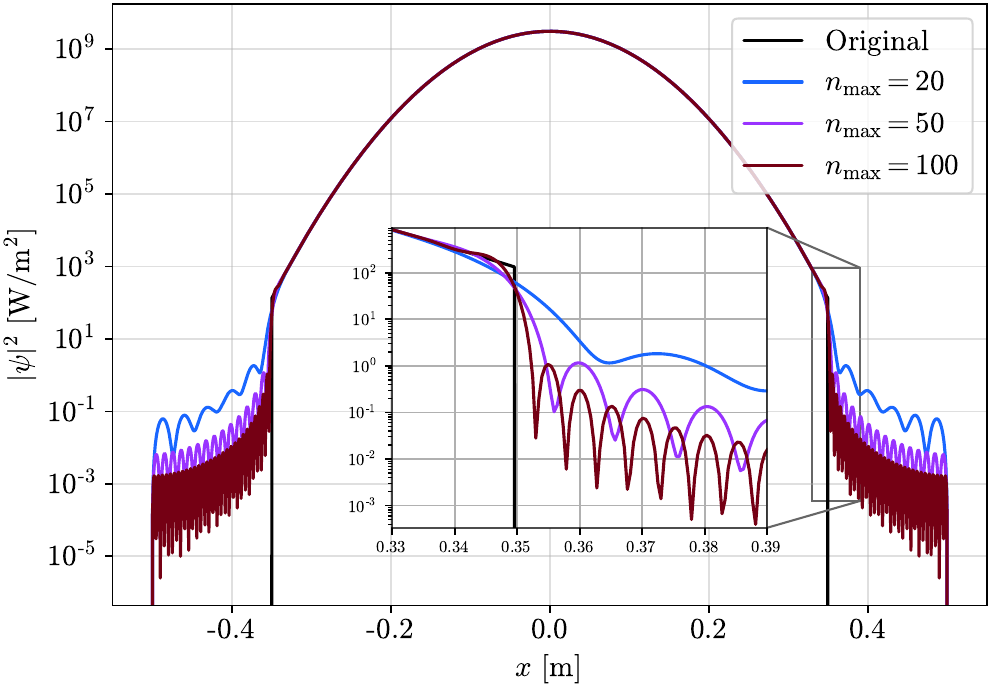}
    \caption{Comparison between the truncated Gaussian beam profile along $y=0$ (black line) and its reconstruction using the waveguide-like modes with azimuthal index $m=0$ and different truncation orders in the radial index, $n_{\max}$.}
    \label{fig:fieldreconstruction}
\end{figure}

Another interesting case is that of reconstructing a Gaussian beam that has been truncated due to a mirror edge or a baffle, reproducing the effect of a component with a finite aperture. Such clipping constitutes one of the sources of scattered light in the interferometer and needs to be captured well by this set of modes. In Fig.~\ref{fig:fieldreconstruction} we show the truncated Gaussian field together with its reconstruction using the waveguide modes. 

In this case, the results show that more modes are required in order to capture a sharper cutoff of the aperture. The quality of the reconstruction improves with the number of radial modes $n_{\max}$ included, converging towards the truncated Gaussian profile as it increases. The sharp cutoff is impossible to be captured perfectly, as that would require an infinite number of modes, but several orders of magnitude of reduction in field intensity are produced over a short distance with $n_{\max} = 100$ (see Fig.~\ref{fig:fieldreconstruction}).

\begin{figure}[htbp]
    \centering
    \includegraphics[width=\linewidth]{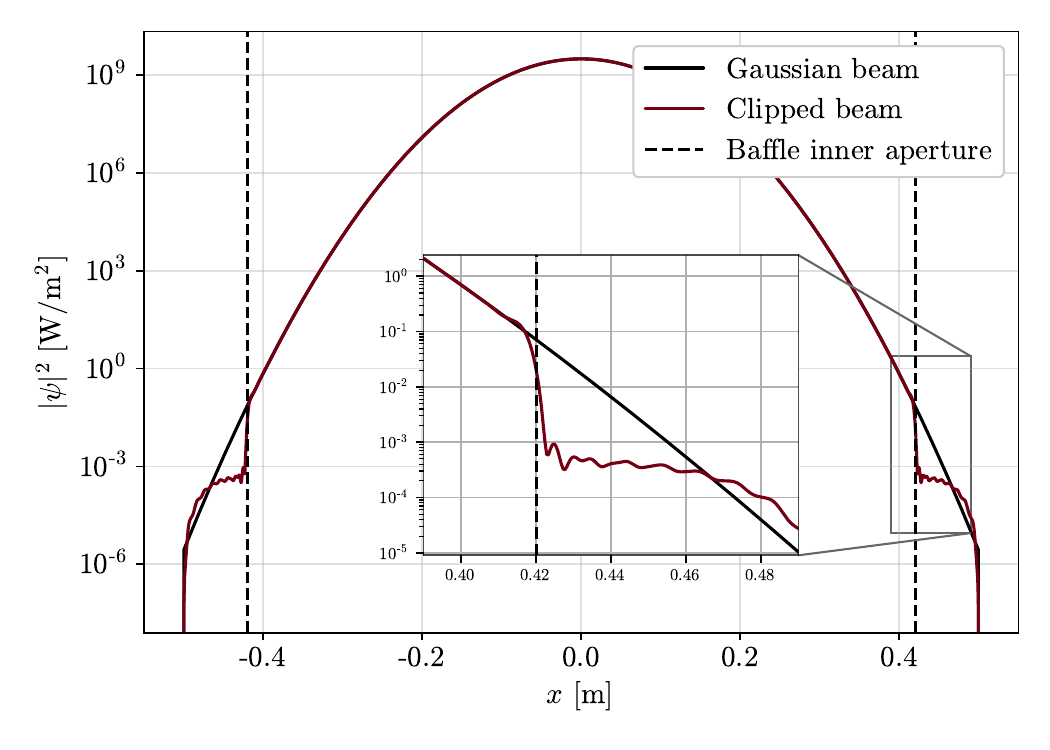}
    \caption{Effect of a baffle aperture on a Gaussian beam reconstructed with the modal expansion formalism. The black curve shows the ideal Gaussian beam intensity profile, while the red curve corresponds to the clipped beam after applying the analytical clipping matrix. Vertical dashed lines mark the baffle's inner aperture.}
    \label{fig:GaussianClipping}
\end{figure}

Finally, we investigate the effect of the presence of a baffle on the propagation of a Gaussian beam using the waveguide mode expansion.
We start with the same Gaussian beam profile reconstructed with $100$ modes, as in Fig.~\ref{fig:fieldreconstruction_Orig}, and then apply the mode-mixing matrix to model the effect of the baffle aperture. The fields before and after clipping are shown in Fig.~\ref{fig:GaussianClipping}. The plot displays the intensity distribution $|\psi|^2$ along the line $y=0$. As expected, the clipped beam follows the Gaussian distribution within the aperture region but exhibits a sharp cutoff of about two orders of magnitude near the baffle edges. Beyond this point, some oscillations arise as a consequence of the truncation, and these dominate the beam profile outside the aperture.

Overall, these results clearly illustrate how the modal expansion method can reproduce both the expected Gaussian behavior inside the baffle aperture and the clipping of the field near the baffle aperture. This further validates the analytical clipping matrix, as is capable of generating the expected sharp cutoff near the baffle edge, like in other modal expansions~\cite{Vajente:22,VajenteCEbaffles}.

\begin{figure*}[htbp]
    \centering
    \includegraphics[width=1.0\linewidth]{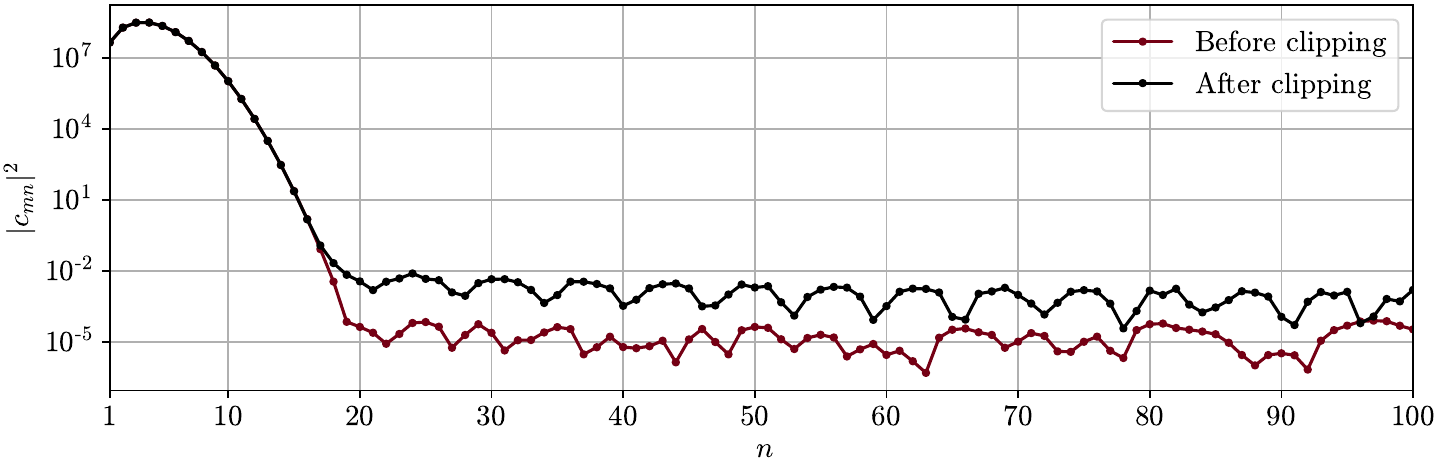}
    \caption{Squared modal expansion coefficients $|c_{mn}|^2$ as a function of the index $n$ for $m=0$. The red curve corresponds to the original Gaussian beam, while the black curve shows the distribution after clipping by the baffle. }
    \label{fig:cmn}
\end{figure*}

In addition to truncating the beam, the baffle also excites higher-order modes, which can later contribute to scattered light noise through diffraction. To quantify this effect, we display the squared modal expansion coefficients $c_{mn}$ as a function of the radial index $n$, with $m=0$ for all cases. The result is shown in Fig.~\ref{fig:cmn}. The figure compares the modal distribution before and after the clipping process. The original Gaussian beam, is strongly dominated by the lowest-order modes and exhibits a rapid decay in power for higher values of $n$. In contrast, after clipping, a significant redistribution of power among higher-order modes is evident. Although the lowest order modes still carry most of the power, the excitation of higher-order contributions is enhanced by about three orders of magnitude. This redistribution highlights how the baffle, while effective in removing part of the Gaussian beam at the aperture, simultaneously generates broader modal content, which in turn can enhance diffraction effects and increase the potential coupling of stray light back into the interferometer.

\subsection{Propagation in the beam tube}
We also benchmark the waveguide propagation developed in this work against the FFT-based propagation of the SIS code~\cite{Romero21}. In both cases we propagate the same injected field, a Gaussian one, through the arm and compare the transverse intensity profile at the end mirror plane, using the parameter set summarized in Tab.~\ref{tab:ParamsTable}. For the waveguide computation we truncate the basis to azimuthal indices up to $m\le 7$ and radial orders up to $n\le 40$, which is sufficient to capture the near-axis structure and the field within the mirror clear aperture.

\begin{figure}[htbp]
    \centering
    \includegraphics[width=1\linewidth]{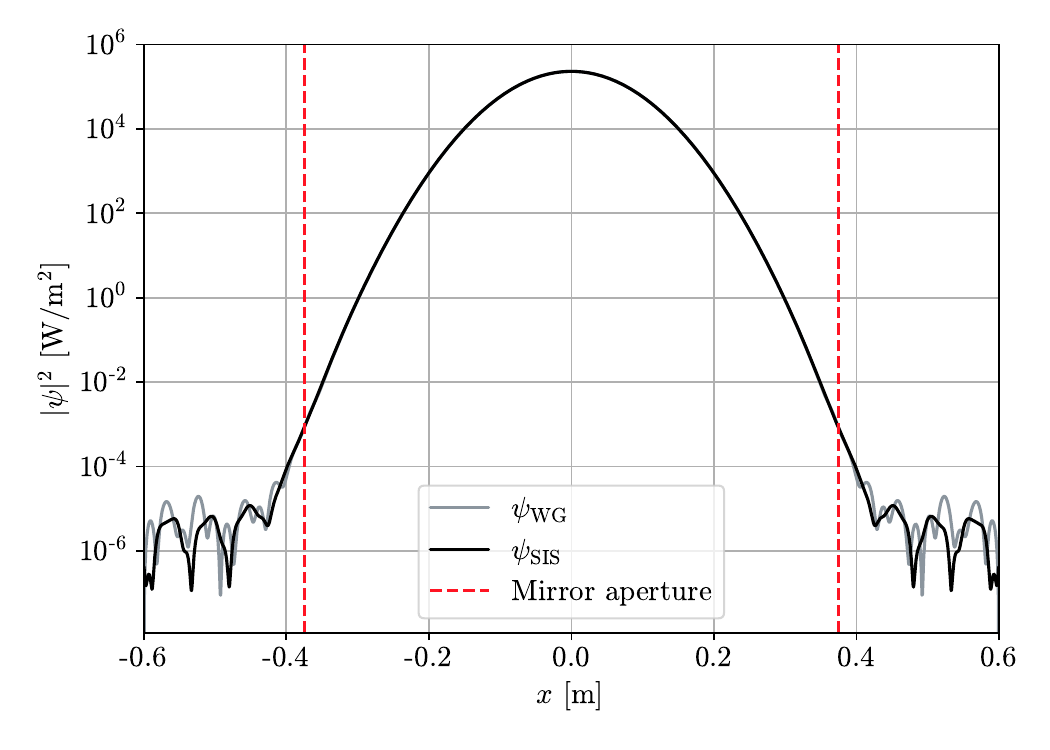}
    \caption{Comparison of the beam intensity after propagation using the waveguide (WG) modal formalism developed and the FFT-based SIS code~\cite{Romero21}. The waveguide result is obtained witho azimuthal indices $m\le 7$ and radial ones $n\le 40$. }
    \label{fig:propagation_comparison}
\end{figure}

Figure~\ref{fig:propagation_comparison} shows a one-dimensional cut of the intensity $|\psi|^2$ along the $x$ axis for the two approaches. Within the mirror aperture (indicated by the vertical dashed lines), the waveguide modes and SIS profiles are in close agreement, demonstrating that a relatively compact modal expansion can accurately reproduce the propagation of the main beam over the length scales of interest. Differences become apparent primarily in the low-amplitude wiggles outside the aperture edge, where the field is most sensitive to the presence of the beam tube itself.

A notable qualitative distinction is the presence of a pronounced oscillatory Airy pattern in the SIS result, characteristic of free-space diffraction from a circular pupil. This behavior is expected from FFT propagation, which evolves the field under the paraxial free-space approximation and therefore produces the familiar ring-like sidelobes caused by the sharp truncation of the mirror aperture. In contrast, the waveguide result does not exhibit the same Airy structure. Since the beam tube is explicitly included through the boundary conditions that define the eigenbasis, the transverse field is confined and its high-angle content is reshaped by the presence of the cylindrical wall. In other words, the free-space picture of the FFT-based code is no longer fully applicable once the beam tube is treated as part of the optical system, and the resulting field outside the clear aperture is modified accordingly. This difference is precisely the effect we aim to capture with the waveguide approach, since the outer part of the field and its spatial structure set the illumination of baffles and other structures and can therefore impact the scattered-light noise.

\subsection{Steady-state field}
We now use the round-trip formalism of Sec.~\ref{sec:steady_state} to compute the steady-state intracavity field in the waveguide basis, and to quantify how a sequence of internal baffles reshapes the cavity resonating field. The optical and geometric parameters are those listed in Tab.~\ref{tab:ParamsTable}. The waveguide expansion is truncated to azimuthal indices up to $m\le 7$ and radial orders up to $n\le 40$ and the steady state is obtained by solving Eq.~\eqref{eq:cIcin} for the modal coefficients $\mathbf{c}_{\mathrm{I}}$.

For the baffled configuration we include a uniform array of $N=200$ centered circular baffles along the arm, with longitudinal positions equally spaced between $1~\mathrm{km}$ and $39.9~\mathrm{km}$ from the ITM. Each baffle is modeled as a circular aperture of radius $R_b$ (see Tab.~\ref{tab:ParamsTable}) through the corresponding clipping matrix $\mathbf{S}(z_i|d_x{=}0)$. This placement is chosen to illustrate the cumulative effect of repeated aperturing on the diffracted field, while avoiding the immediate vicinity of the mirrors.

\begin{figure*}[htbp]
    \centering
    \includegraphics[width=\linewidth]{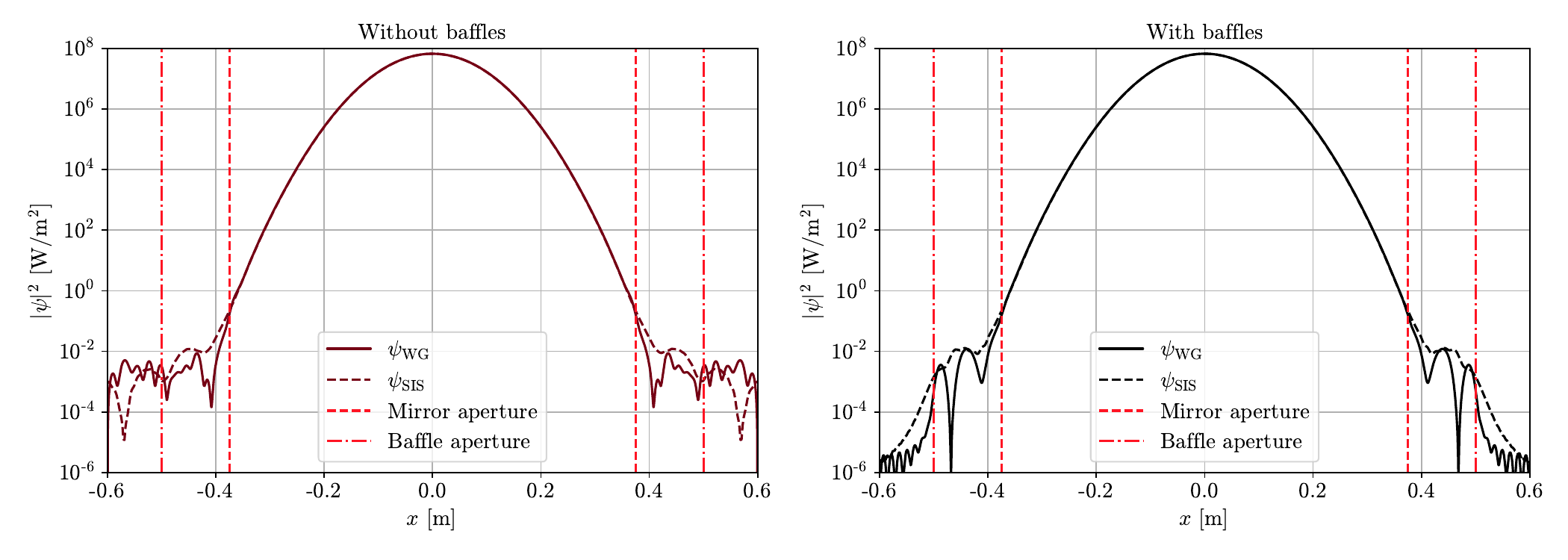}
    \caption{(\textit{Left}) Steady state intensity profile for an unbaffled cavity compared to the SIS results. (\textit{Right}) Same for a cavity with 200 baffles equally spaced between $1$~km and $39.9~$km.}
    \label{fig:steady_state_profile}
\end{figure*}

Figure~\ref{fig:steady_state_profile} shows a one-dimensional cut of the steady-state intensity field impinging into the ITM across the full beam tube diameter, comparing the cavity with and without baffles to the SIS one.
Within the mirror clear aperture (vertical dashed lines), the two solutions are essentially indistinguishable as the steady-state field is dominated by the fundamental mode content and remains well confined inside the optic. The differences become more noticeable in the tails of the beam that fall outside the mirror radius, where the steady-state field is set by the balance between diffraction from the finite mirror aperture and the cavity filtering imposed by propagation and losses. 

In the absence of baffles, the field outside the mirror exhibits a pronounced oscillatory halo extending toward the beam tube radius. This structure is the steady-state manifestation of aperture diffraction as each round trip generates high-spatial-frequency content at the mirror edge, which is then transported along the arm and partially refocused back to the mirror plane. When baffles are introduced, they primarily intercept this diffracted component rather than the main beam. The cumulative effect of many apertures distributed along the arm is therefore to act as a spatial filter attenuating the large-radius content progressively from round trip to round trip and strongly suppressing the intensity approaching the beam tube wall. This is the regime of interest for scattered-light considerations, since it is precisely the tail of field that illuminates baffles and beam tube surfaces and can generate stray-light return paths. 

While the structure obtained with the FFT code is in general compatible with the waveguide modes, some differences arise from the finite number of modes used but also from the effect itself of the presence of the beam tube. In the unbaffled case, this difference becomes more apparent near the edge, where the Airy-like structure of the field is not present. On the other hand, since the baffles reduce the interaction with the beam tube the field presents a more similar structure for the baffled cavity. This comparison indicates that ignoring the presence of the beam tube for a long but baffled cavity in FFT codes is likely a good approximation, which we quantify further in the following sections, from the noise perspective.

\subsection{Noise induced by a displaced baffle}
We next quantify the phase-noise coupling produced by a localized breaking of axial symmetry inside the beam tube. Starting from the nominal baffled cavity considered above and the same WG truncation, $m\le 7$ and $n\le 40$, we introduce a perturbation by transversely displacing a single baffle by an amount $d_x$ along the $x$ direction. All other baffles remain perfectly centered. The displaced baffle is chosen to be the one closest to the middle of the arm, so that the perturbation probes the field after substantial propagation and repeated clipping.

\begin{figure}[htbp]
    \centering
    \includegraphics[width=1\linewidth]{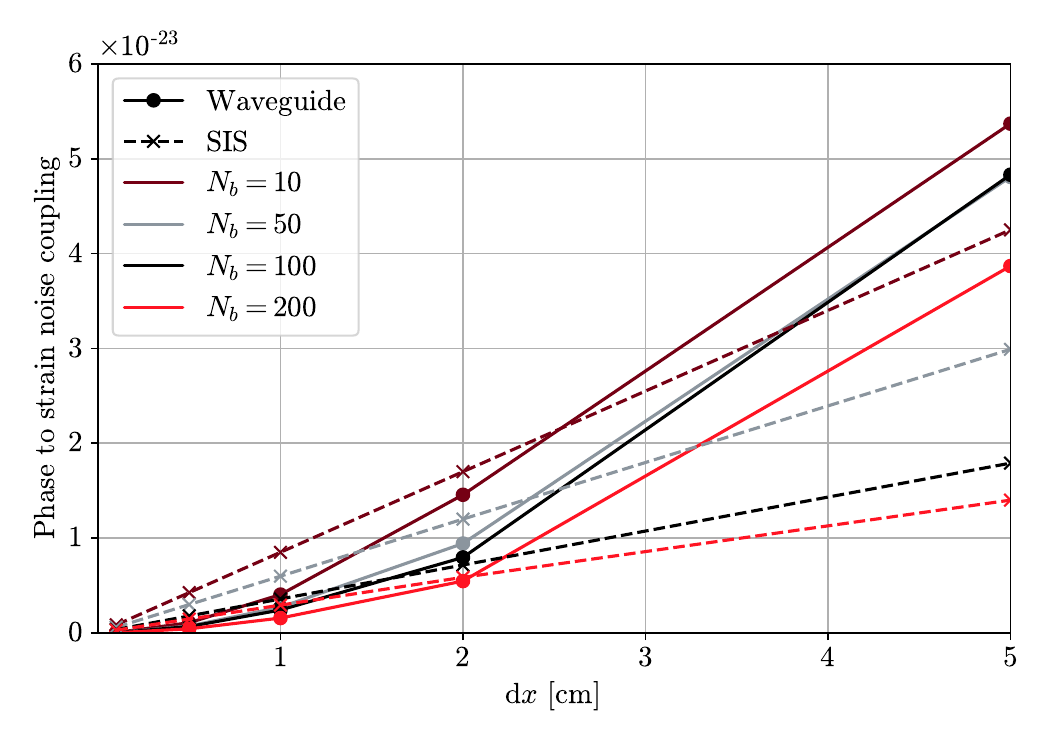}
    \caption{Equivalent strain coupling $h$ induced by transversely displacing the central baffle by $d_x$, shown for several configurations with different total numbers of uniformly spaced baffles along the arm. Solid lines correspond to the waveguide modes approach and dashed lines to SIS.}
    \label{fig:h_dx_baffle}
\end{figure}

For each value of $d_x$ we compute the steady-state intracavity field and extract the induced round-trip phase shift of the fundamental Gaussian component following the prescription of Sec.~\ref{sec:steady_state}. This defines an equivalent strain response $h(d_x)$, which is the quantity displayed in Fig.~\ref{fig:h_dx_baffle}. In addition to the waveguide-mode calculation, we repeat the same procedure using SIS for the same optical configuration and baffle placement, and apply the same phase-extraction prescription for the fundamental component. Figure~\ref{fig:h_dx_baffle} provides a direct comparison between the waveguide and SIS predictions for the same perturbation.

To see the role of baffle spacing, we repeat the same experiment while varying the total number of baffles $N_b$ distributed along the arm. In all cases the baffles are placed uniformly between $z=1~\mathrm{km}$ and $z=39.9~\mathrm{km}$ from the ITM and the displaced element is the central one of the sequence. Decreasing $N_b$ increases the separation between successive baffles and leaves longer sections of beam tube effectively unshielded. This increase of exposed tube allows diffracted and higher-order content generated at the mirrors (and at upstream apertures) to propagate over longer distances before being clipped again.

The trend in Fig.~\ref{fig:h_dx_baffle} follows this physical picture: configurations with fewer baffles (larger spacing) exhibit a larger phase coupling, whereas denser baffle placement suppresses the coupling. In other words, the miscentered baffle predominantly perturbs the weak, large-radius component of the steady-state field, and the magnitude of that component is controlled by how frequently the baffle sequence intercepts the large.radius halo along the arm. The waveguide and SIS results show consistent qualitative behavior and they mostly agree in the small miscentering case, the one most relevant for realistic scenarios of GW detectors. Additionally, it is worth noting that the match improves as the number of baffles increases, because the effect of the beam tube is less prominent. For larger displacements, the two approaches can deviate quantitatively, which is expected as strong miscentering drives enhanced mixing into higher-order spatial content and makes the interaction with the beam tube more relevant. 

These results provide a direct quantification of how baffle spacing and miscentering impacts the noise, but also validates the current SIS approach, as for the relevant regime (densely baffled cavity and small displacements) the results are very much in agreement.

\subsection{Noise of a defect in the tube}
Finally, we explore how a localized imperfection of the beam tube can affect the phase-noise coupling. In practice, small departures from an ideal cylinder can arise from fabrication tolerances, local ovalization, weld seams, or protrusions, and they may effectively introduce a weak additional clipping at large radii. To emulate such a defect within our framework, we insert an extra circular aperture between the two baffles closest to the cavity midpoint. This defect aperture has an inner radius of $R_d=0.58~\mathrm{m}$, slightly smaller than the nominal beam tube radius, and it is transversely displaced by a fixed amount of $d_x=5~\mathrm{mm}$.

\begin{figure}[htbp]
    \centering
    \includegraphics[width=\linewidth]{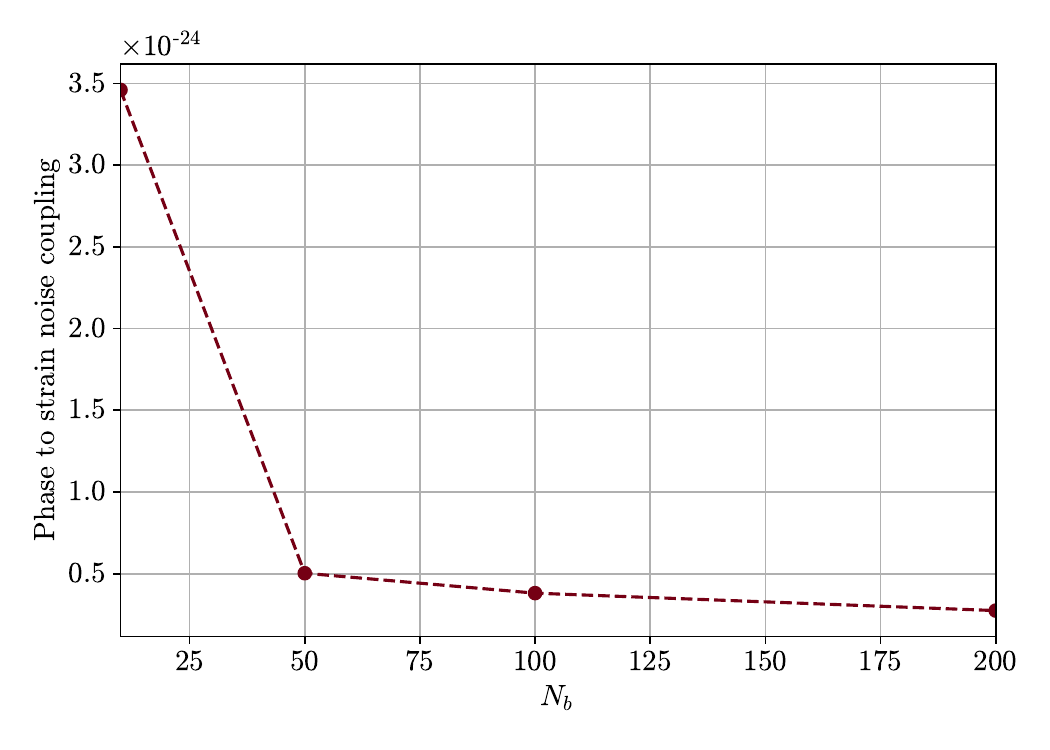}
    \caption{Phase to noise coupling for a localized beam tube defect modeled as an additional off-centered circular aperture of radius $R_d=0.58~\mathrm{m}$ displaced by $d_x=5~\mathrm{mm}$, inserted between the two baffles closest to the cavity midpoint, shown versus the total number of uniformly spaced baffles $N_b$.}
    \label{fig:h_dx_defect}
\end{figure}

We then compute the steady-state field and extract the induced round-trip phase shift of the fundamental Gaussian component using the same phase-noise prescription as in Sec.~\ref{sec:steady_state}. Figure~\ref{fig:h_dx_defect} shows the resulting equivalent strain coupling as a function of the number of baffles $N_b$ used in the arm.

The observed dependence on $N_b$ follows the expected behavior. When fewer baffles are used, the spacing between them increases and larger sections of beam tube are effectively unshielded, allowing more diffracted content to propagate and populate the larger radius. In this regime, even a mild additional clipping near the wall such as the one of the defect considered here can produce a measurable phase perturbation when it is not perfectly centered. As the baffle density increases, the field is progressively filtered by the sequence of apertures, reducing the light incident on the defect plane and correspondingly suppressing the associated phase-noise coupling. This illustrates that the sensitivity to small symmetry-breaking imperfections is not set solely by the mirror apertures, but can also be influenced by localized beam tube defects that interact with the large-radius tail of the intracavity field.

\section{Conclusions}
We have introduced a waveguide-like mode description of the optical field in long Fabry-Perot arm cavities hosted in vacuum beam tubes. This construction incorporates an imposed beam tube boundary condition directly in the simulations, which is relevant in the context of ongoing modeling efforts for next-generation GW detectors and complements standard paraxial free-space propagation tools.

Within this framework, we have derived the modal mixing matrices required to model mirrors and baffles inside the cavity. For the axisymmetric circular-aperture case, we have also obtained a closed-form series representation and validated it against direct numerical integration. Using CE-representative parameters, we have demonstrated that the waveguide basis can accurately reconstruct the type of beams found in GW detectors. In particular, the steady-state field within the mirror clear aperture agrees closely with FFT-based propagation tools, indicating that a moderate truncation in azimuthal and radial indices suffices to capture the main resonant beam over 40~km scales. Differences arise predominantly in the large-radius tail of the field, where the beam tube boundary condition alters the familiar free-space structure. This regime is especially relevant for stray-light studies because it governs the illumination of baffles and beam tube surfaces.

This same formalism has been applied to compute the steady-state intracavity field. For a 40~km cavity with an array of baffles, we find that the circulating field within the mirror aperture remains essentially Gaussian, while the distributed apertures progressively filter the large-radius halo that would otherwise extend toward the tube wall. We show that, in the baffled configurations, the steady-state results are largely compatible with FFT codes, consistent with the beam tube boundary condition having a subdominant impact on the resonant beam in the regime of interest. In contrast, for the unbaffled case the differences are more pronounced.

Finally, we have quantified phase-coupling induced by transverse miscentering of a single baffle and by a localized wall defect. In both cases, the equivalent strain response is larger when baffles are sparse and is suppressed as the baffle density increases, consistent with the interpretation that these perturbations can get filtered by other baffles. This highlights the role of baffles in shielding the tube and mitigating stray-light couplings. We have compared these couplings against SIS and find close agreement in the relevant regime of a densely baffled cavity and small baffle displacements, providing an independent cross-check and supporting the continued use of SIS for such studies.

\begin{acknowledgments}
The authors would like to thank L. Barsotti, A. Kontos and K. Kuns for insightful comments and discussions about this work. This document has received an internal CE DCC number of P2600001 and LIGO DCC number of P2600010. This research has been supported by the National Science Foundation (NSF) awards PHY-2309064, PHY-2308793 and PHY-2308794.
\end{acknowledgments}

\bibliography{ref}

\appendix

\section{Derivation of the waveguide-like complex modes}
\label{sec:newmodes_app}

To account for the beam tube we do not need to use the paraxial approximation and instead we can solve the scalar Helmholtz equation in cylindrical coordinates, which reads as
\begin{equation}\label{eq:HelmCyl}
    \frac{1}{r}\frac{\partial}{\partial r}\left( r\frac{\partial \psi}{\partial r} \right)+\frac{1}{r^2}\frac{\partial^2 \psi}{\partial \phi^2}+\frac{\partial \psi}{\partial z^2}+k^2\psi = 0\, ,
\end{equation}
with Dirichlet and periodic boundary conditions $\psi(r=R,\phi,z) = 0$ and $\psi(r,\phi,z)=\psi(r,\phi+2\pi,z)$. Here, $R$ denotes the radius of the beam tube. This is a standard waveguide-style eigenmode construction and the resulting functions are reminiscent of TE/TM waveguide patterns, but here they should be understood as scalar Helmholtz eigenmodes with an imposed Dirichlet boundary at $r=R$\footnote{Unlike true TE/TM modes, which are solutions of the vector Maxwell equations with electromagnetic boundary conditions and polarization/impedance content, our modes are obtained from the scalar Helmholtz equation for a fixed polarization component with $\psi(R)=0$.}. Assuming an ansatz of the form $\psi = \Psi(r,\phi)e^{-i\beta z}$, Eq.~\eqref{eq:HelmCyl} can be reduced to
\begin{equation}
    \frac{1}{r}\frac{\partial}{\partial r}\left( r\frac{\partial \Psi}{\partial r} \right)+\frac{1}{r^2}\frac{\partial^2 \Psi}{\partial \phi^2}+k_c^2\Psi = 0\, ,
\end{equation}
where $k_c^2=k^2-\beta^2$. This equation also admits a solution in the via separation of variables by further assuming $\Psi(r,\phi) = \rho(r)\Phi(\phi)$.  From this we obtain a system of two independent ordinary differential equations
\begin{equation}\label{eq:Phi}
    \frac{\td^2\Phi}{\td\phi^2}+m^2\Phi=0\,,
\end{equation}
\begin{equation}\label{eq:Rho}
    \frac{r}{\rho}\frac{\td}{\td r}\left( r\frac{\td \rho}{\td r}\right)+k_c^2r^2=m^2\, ,
\end{equation}
with $m$ an integration variable to be determined later to satisfy boundary conditions. 

Equation~\eqref{eq:Phi} can be easily integrated explicitly yielding 
\begin{equation}
    \Phi(\phi) = Ce^{im\phi}\, ,
\end{equation}
with $C\in \mathbb{C}$ being a complex integration constant. Due to the periodicity in $\phi$ demanded by the original boundary conditions, we note that $m$ must be an integer (i.e., $m\in\mathbb{Z}$). 

For the other half of our propagation problem,
Eq.~\eqref{eq:Rho} can be analytically integrated to yield
\begin{equation}
    \rho(r) = D_mJ_{|m|}(k_cr)+E_mY_{|m|}(k_cr)\,,
\end{equation}
with $D_m,E_m\in \mathbb{C}$ as additional complex integration constants and $J_{|m|}$ and $Y_{|m|}$ the Bessel functions of 1st and 2nd kind, respectively. Since $Y_{|m|}(x)$ diverges at $x=0$ we fix $E_m = 0,~\forall m\in\mathbb{Z}$. Imposing the other original boundary condition, $\rho(r=R)=0$, we note how $k_c$ must be related to the roots of the Bessel function of the first kind as 
\begin{equation}
    k_c = \frac{\alpha_{|m|n}}{R}\, ,
\end{equation}
where $\alpha_{|m|n}$ is the $n^{\rm th}$ root of $J_{|m|}$. This also sets the value of the constant $\beta$ related to the propagation in the $z-$direction, as 
\begin{equation}
    \beta_{mn} = \sqrt{k^2-\left(\frac{\alpha_{|m|n}}{R} \right)^2}\, .
\end{equation}

Modes with real values of $\beta_{mn}$ correspond to the propagating ones, and the ones with imaginary $\beta_{mn}$ are evanescent and decay with $z$. Therefore, there is a maximum number of propagating modes, which correspond to the indices $(m,n)$ that satisfy
\begin{equation}
    \alpha_{|m|,n} \leq Rk
\end{equation}

Finally, we can explicitly write the set of orthogonal modes as
\begin{equation}\label{eq:WaveguideModes}
    \psi_{mn}(r,\phi,z) = J_{|m|}\left(\frac{\alpha_{|m|n}r}{R}\right)e^{im\phi}e^{-i\beta_{mn}z}\, .
\end{equation}

Using this new orthogonal basis we can decompose any field (that is defined within the beam tube radius) as a linear combination of infinite waveguide eigenmodes as
\begin{equation}\label{eq:cmn_psi}
    \psi(r,\phi,z) = \sum_{m\in\mathbb{Z}}\sum_{n\in\mathbb{N}} c_{mn}\psi_{mn}(r,\phi,z)\,,
\end{equation}
and each of the coefficients can be computed as 
\begin{equation}
    c_{mn} =  \frac{\langle \psi_{mn},\psi \rangle}{\langle \psi_{mn},\psi_{mn} \rangle} ,
\end{equation}
with the inner product defined as \begin{equation}
    \langle a,b\rangle = \int_Aa^*b~\td A\, ,
\end{equation}
 where $^*$ denotes the complex conjugate and $A$ the transverse area.

In all the examples that we employ in this work, there is the symmetry $\phi\to-\phi$ in the input fields and all the operators. Therefore, in the expansion of Eq.~\ref{eq:cmn_psi}, we have that $c_{-m,n}=c_{m,n}$ and we can pair $\pm m$ terms as
$$
    c_{m,n}e^{im\phi}+c_{-m,n}e^{-im\phi} = 2c_{m,n}\cos(m\phi)\, ,
$$
which is equivalent to setting $m\geq0$ and using the basis 
\begin{equation}\label{eq:WaveguideModes_cos}
    \psi_{mn}(r,\phi,z) = J_{m}\left(\frac{\alpha_{mn}r}{R}\right)\cos(m\phi)e^{-i\beta_{mn}z}\, .
\end{equation}

By doing this, we can keep track of only half of the modes, reducing the computational cost of the numerical experiments.

 \section{Modal-mixing matrices}\label{sec:modal_mixing}

 To work with this new set of modes, we need to first define the operators that can act on the field in this basis. 
The first of them is the propagation operator, which determines how each mode is propagated a distance $\Delta z$. 
From Eq.~\eqref{eq:WaveguideModes} we note that for each mode
\begin{equation}
    \psi_{mn}(r,\phi,z+\Delta z) 
    = e^{-i\beta_{mn}\Delta z}\,\psi_{mn}(r,\phi,z)\, .
\end{equation}
Since there is no mode mixing when the field propagates down the beam tube, the propagation matrix is diagonal:
\begin{equation}
    \mathbf{P}_{mn}(\Delta z) = \mathrm{diag}\!\big(e^{-i\beta_{mn}\Delta z}\big)\, .
\end{equation}

Any other generic operator $\mathcal{Q}$ that acts multiplicatively on the field,
\begin{equation}
    \mathcal{Q}\psi(r,\phi) = Q(r,\phi)\,\psi(r,\phi)\, ,
\end{equation}
has an associated mode--coupling matrix with elements
\begin{equation}\label{eq:Qmnpq}
    \mathbf{Q}_{mn,pq} 
    =  \frac{\langle \psi_{mn},Q\psi_{pq} \rangle }{\langle\psi_{mn},\psi_{mn} \rangle}\, .
\end{equation}

Given a vector of modal coefficients $\{c_{pq}\}$, the action of $\mathcal{Q}$ is to modify these coefficients to
\begin{equation}
    c'_{mn} = \sum_{p,q} \mathbf{Q}_{mn,pq}\, c_{pq}\, .
\end{equation}

If the operator $\mathcal{Q}$ is azimuthally symmetric, meaning that $Q=Q(r)$, the matrix $\mathbf{Q}$ becomes block-diagonal in the azimuthal index and does not couple different $m$ indices. This is especially relevant to increase the computing speed of the mode-coupling matrix in those cases. The remainder of this section introduces the modal-mixing matrices for baffles and mirrors.

\subsection{Mirror and baffle modal-mixing matrices}
\label{sec:mirror_baffle_mixing}

Within the present framework, any thin optical element acting in a  plane transverse to the propagation direction is described by a complex multiplicative profile $Q(r,\phi)$ and an associated mode-coupling matrix $\mathbf{Q}$, whose elements are given by Eq.~\eqref{eq:Qmnpq}. A mirror is therefore modeled as the combination of a scalar complex reflectivity or transmissivity and a spatially dependent mask $Q(r,\phi)$ encoding its finite aperture and any additional phase or amplitude variations. In GW interferometers, the most relevant mirror effects include the finite clear aperture, the spherical radius of curvature, small tilts, and surface height or phase maps~\cite{Romero21}. 

In this work we focus on the impact of finite apertures on the scattered-light field. We assume mirrors to be circular apertures with a clear radius of $R_m$ with uniform amplitude response, so that the corresponding spatial profile reduces to a simple function of
\begin{equation}
   Q(r,\phi) = \Theta(R_m - r)\, ,
\end{equation}
where $\Theta$ denotes the Heaviside step function. More general mirror effects can be incorporated by replacing this mask with the appropriate $Q(r,\phi)$, as described in Ref.~\cite{Romero21}, and computing the associated matrix $\mathbf{Q}$ via Eq.~\eqref{eq:Qmnpq}. The full mirror operators are then obtained by multiplying the modal mask by the scalar reflectivity and transmissivity,
\begin{equation}\label{eq:RQ}
    \mathbf{R} = r\,\mathbf{Q}\,,
\end{equation}
for reflection, and
\begin{equation}\label{eq:TQ}
    \mathbf{T} = t\,\mathbf{Q}\,,
\end{equation}
for transmission. If several mirror-related effects are present simultaneously, their combined action is represented by a single profile $Q(r,\phi)$ that contains all factors, from which the corresponding $\mathbf{Q}$ is constructed once.

Baffles installed inside the beam tube are treated in an entirely analogous way. Such structures are already employed in current detectors and are foreseen as a key stray-light mitigation strategy for next-generation instruments. However, the finite apertures of the baffles introduce diffraction that imprints a characteristic spatial pattern on the circulating field. Preliminary studies for CE and ET indicate that this baffle-induced diffraction can become a dominant source of scattered-light noise in these next-generation instruments if not carefully controlled~\cite{Andres-Carcasona:2023qom,Andres-Carcasona:2025xwq,VajenteCEbaffles}. 

The action of a baffle is described by the same overlap formalism as in Eq.~\eqref{eq:Qmnpq}, with $Q(r,\phi)$ now playing the role of a masking function that equals unity inside the clear aperture and zero elsewhere. For a perfectly circular, centered baffle the mask is again
\begin{equation}\label{eq:Rb_clipping}
   Q(r,\phi) = \Theta(R_b - r)\,,
\end{equation}
with $R_b$ the baffle radius. 

Simple circular apertures for both the baffle and the mirror allow us to derive exact expressions as shown in Sec.~\ref{sec:analyticalMatrices}. For more realistic geometries, such as serrated edges or off-centered baffles, the loss of axial symmetry prevents a simple closed form, and the numerator of Eq.~\eqref{eq:Qmnpq} must be evaluated numerically on a transverse grid. 

In what follows, we denote the clipping matrix generated by a baffle as $\mathbf{S}$, to distinguish it from the mirror-related operator $\mathbf{Q}$.

\subsection{Analytical expression for circular apertures}\label{sec:analyticalMatrices}
Here we derive analytical expressions that allow for fast computation of the mode-mixing matrix and a practical implementation for the analysis of GW experiments. We specifically consider the case of the clipping of baffles or mirrors with a circular finite aperture. The mode-mixing matrix of a circular baffle can be obtained by plugging Eq.~\eqref{eq:Rb_clipping} into Eq.~\eqref{eq:Qmnpq}, which leads to
\begin{equation}
    \mathbf{S}_{mn,pq} = \frac{\displaystyle\int_0^{R_b}r~\td r\int_0^{2\pi}\td \phi~\psi_{mn}^*\psi_{pq} }{\displaystyle\int_0^{R}r~\td r\int_0^{2\pi}\td \phi~\psi_{mn}^*\psi_{mn} }\, .
\end{equation}

The denominator can be explicitly written as 
\begin{multline}
    \int_0^{R}r~\td r\int_0^{2\pi}\td \phi~\psi_{mn}^*\psi_{mn} = \\ \int_0^{R}rJ_m^2\left(\frac{\alpha_{mn}r}{R}\right)~\td r\int_0^{2\pi}\cos^2(m\phi)~\td \phi\, ,
\end{multline}
and if $m=0$, the azimuthal integral yields $2\pi$, while if $m>0$ it equals only $\pi$. To have a compact notation for it, we write the general solution of the azimuthal integral as $\pi(1+\delta_{m0})$, with $\delta_{ab}$ the Kronecker delta. The radial integral yields
\begin{equation}
    \int_0^{R}rJ_m^2\left(\frac{\alpha_{mn}r}{R}\right)~\td r = \frac{1}{2}R^2J_{1+m}^2(\alpha_{mn})\, , 
\end{equation}
so that the normalization factor appearing in the denominator becomes
\begin{equation}
    \int_0^{R}r~\td r\int_0^{2\pi}\td \phi~\psi_{mn}^*\psi_{mn} = \frac{\pi(1+\delta_{m0})}{2}R^2J_{1+m}^2(\alpha_{mn})\, .
\end{equation}

We now shift our attention to the numerator of the coupling matrix. If we expand it with the waveguide modes we obtain 
\begin{multline}
   e^{i(\beta_{mn}-\beta_{pq})z} \int_0^{R_b}rJ_m\left( \frac{\alpha_{mn}r}{R}\right)J_p\left( \frac{\alpha_{pq}r}{R}\right)~\td r\\\times\int_0^{2\pi} \cos(m\phi)\cos(p\phi)~\td\phi\, .
\end{multline}
For $m=p>0$ the result of the azimuthal integral is $\pi$, for $m\neq p$ it is $0$ and for the special case of $m=p=0$ the result is $2\pi$, which can be represented compactly as $\pi(1+\delta_{m0})\delta_{mp}$. The radial integral can also be solved analytically, but not in a straight forward way. Before solving, we need to rewrite the product of two Bessel function of the first kind as an infinite sum, by using the following relation~\cite{watson1995treatise,gradshteyn2014table,math7100978_Bessel}:
\begin{multline}
    J_\mu(ax)J_\nu(bx)=\frac{a^\mu b^\nu x^{\mu+\nu}}{2^{\mu+\nu}\Gamma(\nu+1)}\\ \times \sum_{j\geq 0}\frac{(-1)^j(ax)^{2j}}{j!4^j\Gamma(\mu+j+1)}{}  _2F_1\left[ \begin{smallmatrix}
-j, -\mu - j \\
\nu + 1
\end{smallmatrix} \middle| \frac{b^2}{a^2} \right]\, ,
\end{multline}
with ${}_2F_1$ the hypergeometric function and $\Gamma$ the Gamma function. Each successive term in this expansion is suppressed by the factorial and Gamma function factors in the denominator, while the numerator only scales as $(ax)^{2j}$ plus the contribution from the hypergeometric function, both of which are less than the factorial scaling. 
As a result, the relative contribution of higher-order terms rapidly decreases with increasing $j$. 
For any fixed values of $a$, $b$, and $x$, there exists a finite $j$ beyond which additional terms become negligible compared to the leading contributions. 
Therefore, the series can be safely truncated at moderate $j$ without introducing any appreciable error in the evaluation of $J_\mu(ax)J_\nu(bx)$. 
This property makes the representation particularly convenient for our application.

If we define $a_{mn}\equiv \alpha_{mn}/R$ and $a_{pq}\equiv \alpha_{pq}/R$, the original integrand of the radial integral becomes
\begin{multline}
    rJ_m(a_{mn}r)J_{p}(a_{pq}r)=\frac{a_{mn}^m a_{pq}^p}{2^{m+p}\Gamma(p+1)}\\ \times\sum_{j\geq 0}\frac{(-1)^ja_{mn}^{2j}}{j!4^j\Gamma(m+j+1)}{}_2F_1\left[ \begin{smallmatrix}
-j, -m - j \\
p + 1
\end{smallmatrix} \middle| \frac{a_{pq}^2}{a_{mn}^2} \right]r^{1+2j+m+p}\, .
\end{multline}

This expression only has a dependence in $r$ through an exponential, which allows for an analytical integration, yielding
\begin{equation*}
    \int_{0}^{R_b}r^{1+2j+m+p}~\td r=\frac{R_b^{2+2j+m+p}}{2+2j+m+p}\, .
\end{equation*}

Therefore, the baffle clipping matrix can be written as 
\begin{widetext}
    \begin{equation} \label{eq:Smnpq}
\mathbf{S}_{mn,pq} =\frac{\delta_{mp}e^{i(\beta_{mn}-\beta_{pq})z}}{2^{m+p-1}\Gamma(p+1)R^2J_{1+m}^2(\alpha_{mn})}\sum_{j\geq 0}\frac{(-1)^ja_{mn}^{2j+m}a_{pq}^p}{j!4^j\Gamma(m+j+1)}{}_2F_1\left[ \begin{smallmatrix}
-j, -m - j \\
p + 1
\end{smallmatrix} \middle| \frac{a_{pq}^2}{a_{mn}^2} \right]\frac{R_b^{2+2j+m+p}}{2+2j+m+p}\, .
\end{equation}
\end{widetext}

Since this matrix depends on the $z$-coordinate through a simple exponential term, we will separate it as the element-wise product of two matrices:
\begin{equation}
    \mathbf{S}(z) = \tilde{\mathbf{S}}\odot\mathbf{Z}(z)\, 
\end{equation}
where $\mathbf{Z}_{mn,pq}=e^{i(\beta_{mn}-\beta_{pq})z}$ and $\tilde{\mathbf{S}}$ contains the rest of terms of Eq.~\eqref{eq:Smnpq}.

The mirror mode-mixing matrix if only the finite aperture is considered can be obtained by substituting $R_b$ for $R_m$ in Eq.~\eqref{eq:Smnpq} to obtain $\mathbf{Q}$ and applying the reflectivity and transmissivity coefficients.

\subsection{Miscentered baffle}
We also need to consider the case where a single circular baffle has its center displaced, as the miscentering can be caused by vibrations and will modulate the field introducing a diffraction noise. We assume a displacement by a distance $d_x$ along the $x$--direction with respect to the beam tube axis.  The corresponding multiplicative operator is the characteristic function of a disk centered at $(x,y)=(d_x,0)$, which in polar coordinates is
\begin{equation}
    Q(r,\phi|d_x)
    = \Theta\!\bigl(R_b^2 - r^2 - d_x^2 + 2 r d_x \cos\phi\bigr)\, .
\end{equation}

This operator leads to the modal-mixing mixing matrix of 
\begin{multline}
    \tilde{\mathbf{S}}_{mn,pq}(d_x)
    = 
      \frac{1}{N_{mn}}\int_0^{R} r
      J_m\left(\frac{\alpha_{mn}}{R} r\right)
      J_p\left(\frac{\alpha_{pq}}{R} r\right)
      \\ \times A_{mp}(r|d_x)~\td r\,,
\end{multline}
where $N_{mn}$ is the normalization of mode $(m,n)$ and
\begin{equation}
    A_{mp}(r|d_x)
    = \int_{\Omega(r|d_x)} \cos(m\phi)\cos(p\phi)~\td\phi
\end{equation}
is an angular overlap factor that encodes the non-axisymmetric geometry of the shifted aperture. The term $\Omega(r|d_x)$ indicates the region of integration which in this case becomes a function of the radius $r$. This region is 
\begin{multline}
    \Omega(r|d_x)=\\\begin{cases}
        \phi\in[-\pi,\pi], & r\leq R_b-d_x\,,\\
        \phi\in [-\theta_0(r),\theta_0(r)], & R_b - d_x < r < R_b + d_x\, ,\\
        0, & r\geq R_b+d_x\, ,
    \end{cases}
\end{multline}
where 
\begin{equation}
    \cos\theta_0(r) = \frac{r^2+d_x^2-R_b^2}{2r}\, .
\end{equation}

The integral $A_{mp}(r|d_x)$ can be evaluated analytically and one obtains for radii in $R_b - d_x < r < R_b + d_x$,
\begin{multline}
    A_{mp}(r|d_x) = \\ 
    \begin{cases}
      2\theta_0(r), & m=p=0\,,\\
      \theta_0(r) + \frac{\sin\left(2m\theta_0(r)\right)}{2m}, & m=p>0\,,\\
      \frac{\sin\left[(m-p)\theta_0(r)\right]}{m-p}
      + \frac{\sin\left[(m+p)\theta_0(r)\right]}{m+p}, & m\neq p\,,
    \end{cases}
\end{multline}
 while for $r\leq R_b-d_x$ and $r\geq R_b+d_x$, one recovers $A_{mp}(r|d_x)$ by taking $\theta_0(r)=\pi$ and $\theta_0(r)=0$, respectively.

This expression shows explicitly that, for $d_x\neq 0$, the baffle is no longer azimuthally symmetric and therefore does not conserve the azimuthal index. The matrix $\tilde{\mathbf{S}}(d_x)$ generally couples modes with different $m$ and $p$. At the same time, the reduction of the two-dimensional overlap integral to a one-dimensional radial integral with a known angular factor $A_{mp}(r|d_x)$ allows for an efficient numerical implementation.

\section{Efficient numerical evaluation of the baffle clipping matrix}
\label{app:S_numeric}

Although the calculation of $\mathbf{S}$ admits a closed-form series representation, evaluating it for large mode indices becomes numerically fragile. The coefficients involve factorials and Gamma function evaluations that require extended precision and for higher order modes to reach convergence the sum must be truncated at higher values of $j$. On the other hand, a direct two-dimensional numerical evaluation of Eq.~\eqref{eq:Qmnpq} on a Cartesian grid is also inefficient, as it requires integrating mode-by-mode over the full transverse plane and does not exploit the symmetry of the problem. In this Appendix we describe a fast and robust numerical strategy to compute $\mathbf{S}$ that avoids both limitations.

The baffle mode-mixing matrix for the axisymmetric case is block-diagonal in $m$ and only couples radial indices within each $m$ block. In order to evaluate it efficiently we can use the result for the azimuthal integral, which is exact, and reduce the numerical problem to solving a one dimensional integral as
\begin{multline}
\tilde{\mathbf{S}}_{mn,pq}
= \delta_{mp}
\frac{\pi(1+\delta_{m0})}{N_{mn}}
\\ \times \int_{0}^{R_b} r
J_m\left(\frac{\alpha_{mn}}{R}r\right)
J_m\left(\frac{\alpha_{mq}}{R}r\right)\td r\,.
\label{eq:S_axisymmetric_radial_int}
\end{multline}

We evaluate the remaining radial integral in Eq.~\eqref{eq:S_axisymmetric_radial_int} using a Gauss-Legendre quadrature on a fixed interval. Introducing the dimensionless variables $s=r/R$ and $\xi \equiv R_b/R$, the integral becomes
\begin{multline}
\int_{0}^{R_b} r
J_m\left(\frac{\alpha_{mn}}{R}r\right)
J_m\left(\frac{\alpha_{mq}}{R}r\right)~\td r
\\ =
R^{2}\int_{0}^{\xi} s
J_m(\alpha_{mn} s)
J_m(\alpha_{mq} s) ~\td s\,.
\label{eq:S_dimless}
\end{multline}

We then approximate
\begin{equation}
\int_{0}^{\xi} s f(s) \td s
\simeq
\sum_{k=1}^{K} w_k s_k f(s_k)\,,
\label{eq:GL_basic}
\end{equation}
where $\{s_k,w_k\}_{k=1}^K$ are the Gauss-Legendre nodes and weights mapped from $[-1,1]$ to $[0,\xi]$, and $K$ is the number of quadrature nodes.

A key point is that the same node set $\{s_k\}$ can be reused for all matrix elements and for all $m$ blocks, since the integration domain and the weight $s$ are common. This turns the computation of $\tilde{\mathbf{S}}$ into the evaluation of Bessel functions on a fixed grid, followed by linear-algebra operations.

For a fixed azimuthal index $m$, define the matrix
\begin{equation}
\Phi^{(m)}_{n k} \equiv J_m(\alpha_{mn}s_k)\sqrt{w_ks_k}\,,
\label{eq:Phi_def}
\end{equation}
where $n$ labels radial order and $k$ labels quadrature nodes. With this definition, the quadrature approximation of Eq.~\eqref{eq:S_dimless} becomes
\begin{align*}
R^{2}&\int_{0}^{\xi} s J_m(\alpha_{mn}s) J_m(\alpha_{mq}s) \td s
\\ &\simeq
R^{2}\sum_{k=1}^{K}
\Bigl[J_m(\alpha_{mn}s_k)\sqrt{w_k s_k}\Bigr]
\Bigl[J_m(\alpha_{mq}s_k)\sqrt{w_k s_k}\Bigr]
\\ &=
R^{2}(\Phi^{(m)}\Phi^{(m)\,T})_{nq}\,.
\label{eq:Gram}
\end{align*}
Therefore, each $m$ block of $\mathbf{S}$ is obtained as a Gram matrix,
\begin{equation}
\tilde{\mathbf{S}}^{(m)} \simeq
\pi\,(1+\delta_{m0})\,
\left(\mathbf{D}^{(m)}\right)^{-1}\,
\Bigl[R^{2}\,\Phi^{(m)}\Phi^{(m)\,T}\Bigr]\,.
\label{eq:S_block_final}
\end{equation}

where $\mathbf{D}^{(m)}$ is a diagonal matrix containing the normalizations $N_{mn}$ for the modes in the block, and the division by $N_{mn}$ is applied row-wise. The full matrix is assembled by placing the blocks along the diagonal,
\begin{equation}
\tilde{\mathbf{S}} = \mathrm{diag}\bigl(\tilde{\mathbf{S}}^{(m_1)}, \tilde{\mathbf{S}}^{(m_2)}, \ldots \bigr)\,,
\label{eq:S_blockdiag}
\end{equation}
reflecting the selection rule $\delta_{mp}$.

This approach has several practical advantages. First, it avoids the factorial and Gamma growth present in the analytical series representation, relying instead on stable evaluations of Bessel functions at moderate arguments and a well-conditioned quadrature rule. Second, it reduces the original two-dimensional overlap integral to a one-dimensional radial quadrature, and it avoids mode-by-mode Cartesian integrations. Third, it exploits the block-diagonal structure in $m$, so only the non-zero blocks are computed.

Finally, the dominant operations are evaluating the Bessel functions $J_m(\alpha_{mn}s_k)$ on a fixed grid and forming the Gram matrices $\Phi^{(m)}\Phi^{(m)\,T}$. For a given $m$ block with $N_m$ radial modes and $K$ quadrature nodes, the cost scales as $\mathcal{O}(N_m K)$ for the Bessel evaluation and $\mathcal{O}(N_m^2 K)$ for the Gram product. In practice this yields a substantial speed-up compared to a direct Cartesian integration, especially when $\tilde{\mathbf{S}}$ must be computed repeatedly for different parameter sets.

The numerical procedure described here is the one implemented in the code used throughout this work to construct the baffle clipping matrix $\tilde{\mathbf{S}}$ as it yields faster results than the analytical expression for higher order modes. Nonetheless, we have used the important knowledge derived from the analytical derivation, specially the block diagonal structure enforced by the azimuthal integral solution.

\end{document}